\documentclass[10pt, conference]{IEEEtran}
\usepackage{graphicx}
\usepackage{subfigure}

\ifCLASSINFOpdf
\else
\fi
\usepackage[cmex10]{amsmath}

\usepackage {amsmath}
\usepackage {amssymb}
\usepackage {latexsym}
\usepackage {graphicx}
\usepackage {cite}
\usepackage {subfigure}
\usepackage {url}
\usepackage {stfloats}
\usepackage {mathrsfs}
\usepackage{setspace}

\IEEEoverridecommandlockouts

\newcommand{\ls}[1]
    {\dimen0=\fontdimen6\the\font
     \lineskip=#1\dimen0
     \advance\lineskip.5\fontdimen5\the\font
     \advance\lineskip-\dimen0
     \lineskiplimit=.9\lineskip
     \baselineskip=\lineskip
     \advance\baselineskip\dimen0
     \normallineskip\lineskip
     \normallineskiplimit\lineskiplimit
     \normalbaselineskip\baselineskip
     \ignorespaces
}

\begin{document}
\title{A Multi-Channel Diversity Based MAC Protocol for Power-Constrained Cognitive Ad Hoc Networks}

\author{\IEEEauthorblockN{Yichen Wang, Pinyi Ren, Qinghe Du, and Chao Zhang}
\IEEEauthorblockA{School of Electronic and Information Engineering\\
Xi'an Jiaotong University, P. R. China\\
E-mail: \{wangyichen.0819@stu.xjtu.edu.cn, \{pyren, duqinghe, and
chaozhang\}@mail.xjtu.edu.cn\}}
\thanks{The research reported in this paper (correspondence author: Pinyi Ren) was supported in part by the National Natural Science Foundation of
China under Grant No. 60832007 and the National Hi-Tech Research and
Development Plan of China under Grant No. 2009AA011801. } }

\maketitle

\begin{abstract}
One of the major challenges in the medium access control (MAC)
protocol design over cognitive Ad Hoc networks (CAHNs) is how to
efficiently utilize multiple opportunistic channels, which vary
dynamically and are subject to limited power resources. To overcome
this challenge, in this paper we first propose a novel diversity
technology called \emph{Multi-Channel Diversity} (MCD), allowing
each secondary node to use multiple channels simultaneously with
only one radio per node under the upperbounded power. Using the
proposed MCD, we develop a MCD based MAC (MCD-MAC) protocol, which
can efficiently utilize available channel resources through joint
power-channel allocation. Particularly, we convert the joint
power-channel allocation to the Multiple-Choice Knapsack Problem,
such that we can obtain the optimal transmission strategy to
maximize the network throughput through dynamic programming.
Simulation results show that our proposed MCD-MAC protocol can
significantly increase the network throughput as compared to the
existing protocols.

\end{abstract}


% \fi
\IEEEpeerreviewmaketitle

\ls{0.89}
\section{Introduction}

\IEEEPARstart{C}{ognitive radio} is a promising yet challenging
technology to solve wireless-spectrum underutilization problem
caused by the traditional static spectrum allocation strategy [1].
Built upon the cognitive radio (CR) technology, cognitive Ad Hoc
networks (CAHNs), playing a critically important role in future
wireless networks, have attracted a great deal of research
attention. In CAHNs, MAC protocols are responsible for dynamically
accessing opportunistic channel for packet transmission.
Correspondingly, one of the most important targets in cognitive MAC
protocol design is how to efficiently use available channels and
limited power budget to increase the network throughput and
guarantee QoS, where QoS provisioning is one of the most important
issues in next-generation wireless
networks~[1-12],\,\cite{Q-Du-magzine,X-Zhang,Jia-Tang-relay,Xi-Zhang3,J-Tang-twc-jun-2008,J-Tang}.

Diversity technologies are widely used to improve the throughput of
Ad Hoc networks. Three main diversity technologies have been
extensively investigated in recent research, namely Channel
Diversity [2], Link Diversity [3], and Multi-Radio Diversity [4].
However, some drawbacks in these diversity technologies prevent the
network throughput from being further improved. In particular, since
channel diversity and link diversity only use one channel for packet
transmissions, they cannot sufficiently utilize available channel
resources. Although multi-radio diversity can use multiple channels
simultaneously, mobile nodes need to be equipped with multiple
radios, increasing the implementation cost and power consumption. In
addition to these diversity technologies, Game Theory [5-7] and
Water-filling Algorithm [8-9] have been also applied into CAHNs for
resource allocation. These two types of methods aim at identifying
the optimal transmission rate to maximize the network throughput. In
these two methods, the transmission rate can vary continuously,
which, however, cannot be implemented in practical systems. For
example, IEEE 802.11b only supports four different transmission
rates, which are equal to 1, 2, 5.5, and 11~Mbps, respectively, and
802.11a only supports eight rates ranging from 6 to 54~Mbps.
Moreover, these two methods yield high computational complexity,
making them less attractive for practical CAHNs.

In order to efficiently use available channel resources, which vary
dynamically with time, under limited power resources, we first
propose a novel diversity technology called \emph{Multi-Channel
Diversity} (MCD). Our proposed diversity technology is based on the
software-defined radio (SDR), which needs only one radio per
secondary node and allows secondary nodes to use multiple channels
simultaneously through channel aggregation. Using the MCD, we
develop a MAC protocol, called MCD based MAC (MCD-MAC) in
power-constrained cognitive Ad Hoc networks. In our proposed
protocol, control packets are exchanged on the common control
channel (CCC). Each node pair that successfully finishes the control
packets exchange will perform joint power-channel allocation and
continuously transmit multiple data packets. In order to maximize
the network throughput, we convert the joint power-channel
allocation to the Multiple-Choice Knapsack Problem and obtain the
optimal allocation scheme through dynamic programming. In addition,
we also design a mechanism to guarantee the \emph{transmission-time
fairness} for different node pairs.

The rest of this paper is organized as follows.
Section~\ref{section_system_model} presents the network model.
Section~\ref{MCD-MAC_Protocol} defines the Multi-Channel Diversity
and proposes the MCD-MAC protocol.
Section~\ref{Performance_Analysis} theoretically analyzes the
performance of MCD-MAC. Simulation results are given in
Section~\ref{Simulation_Results}. The paper concludes with
Section~\ref{Conclusion}.

\section{Network Model}
\label{section_system_model}

\begin{figure}[t]
\centering
\includegraphics[width=8cm]{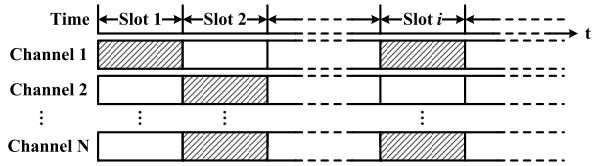}
\caption{The ON/OFF channel usage model}\vspace{-15pt} \label{fig_1}
\end{figure}

Suppose that the CAHN contains one Common Control Channel (CCC) and
\emph{N} Data Channels (DCs). The CCC with central frequency $f_0$
belongs to the CAHN, which is used to exchange control packets. The
DCs with central frequencies $\left\{ {f_1 , \cdots ,f_N } \right\}$
are licensed to Primary Users (PUs). The CAHN can only use those
opportunistic DCs which are not occupied by PUs.

We use the ON/OFF model to characterize the channel-usage statuses
of PUs, as shown in Fig. 1. Specifically, DCs are divided into
synchronized Channel Slots (CSs). The shaded rectangle areas denote
the channels used by PUs, and the white rectangle areas represent
the available channels which can be used by the CAHN. At the
beginning of each CS, PUs independently select which DCs they will
use and the channel usage status of PUs remain unchanged in each CS.

In the CAHN, control packets are transmitted with basic rate
$R_{\rm{{basic}}}$. The set of data-packet transmission rate is
denoted by $R \buildrel \Delta \over = \left\{ {R_1 ,R_2 , \cdots
,R_Q } \right\}$, where $Q$ is the cardinality of $R$ and $R_1  <
R_2 < \cdots < R_Q $. The corresponding set of Signal-to-Noise Ratio
(SNR) is denoted by $SNR = \left\{ {SNR_1 , SNR_2 , \cdots , SNR_Q }
\right\}$, where $SNR_1 < SNR_2 < \cdots < SINR_Q $. The maximum
transmit power for each node is denoted by $P_{\rm{{max}}}$. The
radio propagation model between two nodes follows the two-ray
model~\cite{F-Chen}. Then, the received power is given by
\begin{equation}
P_r (d) = P_t G_t G_r h_t^2 h_r^2 /(d^4 L),
\end{equation}
where $G_t$ and $G_r$ are the gains of transmitter and receiver
antennas, $h_t$ and $h_r$ are the height of transmitter and receiver
antennas, $L$ is the system loss factor, and $d$ is the distance
between transmitter and receiver. Moreover, each node is equipped
with two radios, called control radio and data radio, respectively.
The control radio is devoted to operating on the CCC for control
packet exchanges. The data radio works on the data channels for
sensing, transmitting and receiving. The data radio is based on the
software-defined radio (SDR) so that it can realize channel
aggregation and use multiple channels with different transmit power
simultaneously.

%Moreover, each node is equipped with one radio, which is used for
%sensing, transmitting and receiving, but only one function can be
%implemented at the same time. The length of RTS, CTS, RES, DATA, and
%ACK packets are $L_{rts}$, $L_{cts}$, $L_{res}$, $L_{data}$, and
%$L_{ack}$ respectively. The transmission time of RTS, CTS, RES,
%DATA, and ACK packets are $T_{rts}$, $T_{cts}$, $T_{res}$,
%$T_{data}$, and $T_{ack}$ respectively. Furthermore, the duration of
%Short Inter-Frame Space (SIFS), Distributed Coordination Function
%Inter-Frame Space (DIFS), and one time slot in the CAHN are
%$T_{sifs}$, $T_{difs}$, and $\delta$.

\section{MCD-MAC Protocol}
\label{MCD-MAC_Protocol}

\subsection{Multi-Channel Diversity}

In recent researches, there are three main diversity technologies,
which are Channel Diversity, Link Diversity, and Multi-Radio
Diversity respectively. Although above diversity technologies can
efficiently improve the network throughput, there are still some
drawbacks in them. For channel diversity and link diversity, as only
one channel is allowed for packet transmission, they cannot
sufficiently utilize available channel resources, which will prevent
the throughput from further increasing. For multi-radio diversity,
several channels can be used simultaneously, but nodes in the
network have to be equipped with several radios, which will increase
the cost payment and power consumption.

Therefore, in order to efficiently use available channel resources
and not to bring additional cost, we introduce a novel diversity
technology named Multi-Channel Diversity. The proposed diversity can
be described as: there are several available channels between source
and destination, the node pair first decides the channels and
corresponding power that they will use through exchanging control
packets, then the source continuously transmits multiple packets to
the destination according to the allocation result of channel and
power with only one radio and limited transmitting power.

\begin{figure*}[t] \centering
\includegraphics[width=18cm]{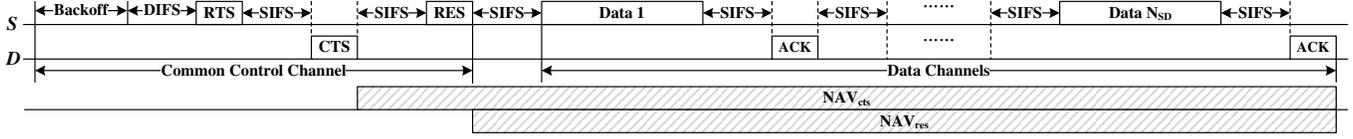}
\vspace{-8pt}\caption{Transmission process of MCD-MAC}\vspace{-17pt}
\label{fig_2}
\end{figure*}

\subsection{Protocol Description}

The MCD-MAC protocol divides each CS into two parts, named Sensing
Period and Data Transmission Period. In sensing period, nodes sense
all DCs to determine opportunistic DCs that PUs do not use in
current CS. In data transmission period, nodes compete for these
opportunistic DCs to transmit data packets through exchanging
control packets on CCC.

Each node maintains one Data Channel Usage List (DCUL). The list
records a item for each DC, and each item contains five parts, which
are ``Channel Number \emph{k}'', ``PU Status'', ``Neighbor Status'',
``Suffered Interference $P_{\inf } \left( k \right)$'', and
``Maximal Allowed Transmitting Power $P_{{\rm{max - s}}} (k)$. ``PU
Status'' and ``Neighbor Status'' represent whether the \emph{k}th DC
is occupied by PUs or neighbor nodes.

In the MCD-MAC protocol, node pairs compete for the opportunistic
DCs through three control packets, which are RTS, CTS, and RES. If
one node pair won the competition, then the two nodes determine the
DCs that they will use for packet transmission, and finish their
packet transmission on those DCs. Suppose the source and destination
nodes are \emph{S} and \emph{D} respectively. The transmission
process is shown in Fig. 2.

\subsubsection{Sending RTS} \emph{S} first overhears on CCC. If the CCC
is busy, then \emph{S} chooses a backoff time and defer its
transmission. If the CCC is idle for a duration of one DIFS after
the backoff time, then a RTS packet that contains the DCUL of
\emph{S} will be sent to \emph{D}.

\subsubsection{Sending CTS} If \emph{D} successfully receives the RTS,
then it compares its DCUL with that of \emph{S}. If common available
DCs, which mean such DCs are not occupied by PUs for both \emph{S}
and \emph{D}, exist in the two DCULs, \emph{D} first computes the
channel gain between itself and S on CCC through (2):
\begin{equation}
h_{SD}^0  = {{P_{\rm{r}}^{{\rm{RTS}}} } \mathord{\left/
 {\vphantom {{P_{\rm{r}}^{{\rm{RTS}}} } {P_{\max } }}} \right.
 \kern-\nulldelimiterspace} {P_{\max } }},
\end{equation}
where $P_{\rm{r}}^{{\rm{RTS}}} $ is the receiving power of the RTS.

Suppose there are \emph{M} common available channels with central
frequencies $\left\{ {f_1 , \cdots ,f_M } \right\}$ between \emph{S}
and \emph{D}, then channel gains on these channels can be acquired
through (3):
\begin{equation}
h_{SD}^m  = h_{SD}^0  \times \left( {{{f_0 } \mathord{\left/
 {\vphantom {{f_0 } {f_m }}} \right.
 \kern-\nulldelimiterspace} {f_m }}} \right)^4 ,~~~ m = 1, \cdots
 ,M.
\end{equation}
Then \emph{D} processes the joint power-channel allocation based on
the information of the two DCULs, and decides the number of data
packets that can be transmitted in the following transmission
process. Finally, a CTS, which contains the information of the
result of power-channel allocation and the number of data packets
for transmitting, is sent to \emph{S}.

\subsubsection{Sending RES} If \emph{S} successfully receives the CTS,
then a RES that contains the same content with the CTS is sent.

\subsubsection{Data packets transmission} After exchanging control
packets, \emph{S} and \emph{D} switch to the corresponding DCs, and
finish their packet transmission.

Furthermore, nodes that overheard CTS or RES packets also need to
modify the information contained in DCULs.

Suppose \emph{S} and \emph{D} will transmit $N_{SD}$ packets with
transmission rate $R_{SD}$, and the power allocation is $\left\{
{P_{SD}^1 , \cdots ,P_{SD}^M } \right\}$. If node \emph{I} overheard
the CTS sent by \emph{D}, \emph{I} first computes the channel gains
$h_{ID}^0 $ and $\left\{ {h_{ID}^1 , \cdots ,h_{ID}^M } \right\}$.
Then for each channel $m$ ($m=1, \cdots ,M$), node \emph{I} computes
the interference caused by the ACK packets transmission and updates
the total interference according to (4):
\begin{equation}
\left\{ \begin{array}{l}
 P_{\inf }^D (m) = P_{SD}^m  \cdot h_{ID}^m  \vspace{3pt}
 \\
 P_{\inf } (m) = P_{\inf } (m) + P_{\inf }^D (m). \\
 \end{array} \right.
\end{equation}
Then the maximum allowed power can be got through (5):
\begin{equation}
P_{\max  - s} \left( m \right) = {{P_{\min }^{\inf } }
\mathord{\left/
 {\vphantom {{P_{\min }^{\inf } } {h_{ID}^m }}} \right.
 \kern-\nulldelimiterspace} {h_{ID}^m }},\;\;m = 1, \cdots ,M,
\end{equation}
where $P_{min}^{inf}$ is the maximum interference power that
neighbor nodes can tolerate. Finally, node \emph{I} updates the
Network Allocation Vector (NAV) of the CTS through (6):
\begin{equation}
NAV_{cts}  \!=\! \frac{{L_{res} }}{{R_{basic} }}\! +\! N_{SD}
\frac{{L_{data} \! + \!L_{ack} }}{{R_{SD} }}\! + \!(2N_{SD} \! +\!
1)T_{sifs}.
\end{equation}
The same way can be used for those nodes that overheard the RES
packet to update correlative information.\vspace{-3pt}

\subsection{Transmission-Time Fairness}

As the MCD-MAC protocol allows node \emph{S} continuously transmits
several packets to node \emph{D}, we must avoid that one node pair
occupies DCs for a long time, which will harm the fairness of the
protocol. Therefore, we use (7) to define the maximum transmission
time, which represents the time that nodes transmit one data packet
with basic rate on CCC and is also used in [2]
\begin{equation}
T_{\max }  = {{L_{{\rm{data}}} } \mathord{\left/
 {\vphantom {{L_{{\rm{data}}} } {R_{{\rm{basic}}} }}} \right.
 \kern-\nulldelimiterspace} {R_{{\rm{basic}}} }}.
\end{equation}

Besides, as varying channel gains will affect power-channel
allocation, the transmission time cannot exceed the coherent time of
DCs. Therefore, the transmission time should satisfy the following
inequality:
\begin{equation}
T_{SD}  \le \min \left\{ {CT(f_1 ), \cdots ,CT(f_M )} \right\} =
CT_{\min },
\end{equation}
where $CT\left( {f_m } \right)$ is the coherent time of DC with
central frequency $f_m$. Therefore, the transmission time $T_{SD}$
has to satisfy:
\begin{equation}
T_{SD}  \le \min \left\{ {CT_{\min } \;,\;\;T_{\max } } \right\}.
\end{equation}
As the transmission time can be acquired as (10), the number of data
packets $N_{SD}$ that node pair is allowed to transmit must hold the
condition shown in (11):
\begin{equation}
T_{SD}  = (2N_{SD}  - 1)T_{sifs}  + N_{SD} (L_{data}  + L_{ack}
)/R_{SD},
\end{equation}
\begin{equation}
N_{SD}  \le \left\lfloor {\frac{{R_{SD} \left( {\min \{ CT_{\min }
,\;\;T_{\max } \} } \right) + T_{{\rm{sifs}}} }}{{L_{{\rm{data}}}  +
L_{{\rm{ack}}}  + 2T_{{\rm{sifs}}} R_{SD} }}} \right\rfloor,
\end{equation}
where $L_{{\rm{data}}}$ and $L_{{\rm{ack}}}$ are the lengths of data
and ACK packets, $T_{{\rm{sifs}}}$ are the the duration of Short
Inter-Frame Space (SIFS), and $\left\lfloor \cdot \right\rfloor$ is
the floor function.

\subsection{Joint Power-Channel Allocation}

In this section, we first convert the joint power/channel allocation
to the Multiple-Choice Knapsack Problem, then obtain the optimal
allocation scheme through dynamic programming.

Suppose the number of common available channels for \emph{S} and
\emph{D} is \emph{M} and the corresponding channel gains are
$\left\{ {h_{SD}^1 , \cdots ,h_{SD}^M } \right\}$. Construct the
matrix of available transmission rates:
\begin{equation}
{\bf{\overline R}} = \left[ {{\bf{\overline R}}^1 , \cdots
,{\bf{\overline R}}^M } \right]^T,
\end{equation}
where ${\bf{\overline R}}^m  = \left[ {\overline R^{m,1} , \cdots
,\overline R^{m,Q} } \right]$ is available transmission rate vector
of the \emph{m}th DC and satisfies $ R^{m,q} = R_q$, $\forall m \in
\{ 1, \cdots , M \}$ and $q \in \{ 1, \cdots , Q \}$. Construct the
matrix of transmit power:
\begin{equation}
{\bf{\overline P}}_{SD}  = \left[ {{\bf{\overline P}}_{SD}^1 ,
\cdots ,{\bf{\overline P}}_{SD}^M } \right]^T,
\end{equation}
where ${\bf{\overline   P}}_{SD}^m  = [ {\overline P_{SD}^{m,1} ,
\cdots ,\overline P_{SD}^{m,Q} } ]$ is the transmit power vector
corresponding to ${\bf{\overline R}}^m$. For $\forall m \in \{ 1,
\cdots , M \}$ and $q \in \{ 1, \cdots , Q \}$, $\overline
P_{SD}^{m,q} $ is the transmit power that \emph{S} sends data
packets with rate $R_q$ on the \emph{m}th DC and can be calculated
as:
\begin{equation}
\overline P_{SD}^{m,q}  = {{SINR_q  \cdot \left[ {P_n  + P_{\inf }
\left( m \right)} \right]} \mathord{\left/
 {\vphantom {{SINR_q  \cdot \left[ {P_n  + P_{\inf } \left( m \right)} \right]} {h_{SD}^m }}} \right.
 \kern-\nulldelimiterspace} {h_{SD}^m }},
\end{equation}
where $P_n$ is the noise power.

\textbf{Problem Description:} Source \emph{S} selects transmit power
on the \emph{M} DCs from vectors ${\bf{\overline P}}_{SD}^1, \cdots,
{\bf{\overline P}}_{SD}^M$, and at most one power can be chosen from
each vector. The optimization objective is to maximize the
transmission rate $R_{SD}$ under the constraints that the total
transmit power and the power used for the \emph{m}th DC are no
larger than $P_{{\rm{max}}}$ and the maximum allowed transmit power
$P_{{\rm{max - s}}} (m)$, respectively.

The above problem is the Multiple-Choice Knapsack Problem and its
mathematical description is
\begin{equation}
\begin{array}{l}
 \mathop {\max }\limits_{\left\{ {x^{m,q} } \right\}} \sum\limits_{m \in \mathcal {M}} {\sum\limits_{q \in \mathcal {Q}_m} {R^{m,q} x^{m,q} } } , \vspace{3pt}
 \\
 {\rm{s}}{\rm{.t}}{\rm{.}}\;\left\{ \begin{array}{l}
 \sum\limits_{m \in \mathcal {M}} {\sum\limits_{q \in \mathcal {Q}_m} {\overline P_{SD}^{m,q}   x^{m,q} } }  \le P_{\max } ,  \vspace{3pt}
 \\
 \overline P_{SD}^{m,q}  \le P_{{\rm{max - s}}} (m),\;m \in \mathcal {M},\;q \in \mathcal {Q}_m,  \vspace{3pt}
 \\
 \sum\limits_{q \in \mathcal {Q}_m} {x^{m,q}  \le 1,\;} m \in \mathcal {M},  \vspace{3pt}
 \\
 x^{m,q}  \in \{ 0,\;1\} ,\;m \in \mathcal {M},\;q \in \mathcal {Q}_m,  \vspace{3pt}
 \end{array} \right. \\
 \end{array}
\end{equation}
where $\mathcal {M}=\{1,\cdots,M\}$ and $\mathcal
{Q}_m=\{1,\cdots,Q\}$, $\forall m \in \mathcal {M}$. For $ \forall i
\in \mathcal {Q}_m $, if $ \overline P_{SD}^{m,i} > P_{{\rm{max}}}
$, then $ i $ is said to be \emph{IP-infeasible} and can be deleted
from $\mathcal {Q}_m$. For $\forall i, j \in \mathcal {Q}_m $, $i$
is said to \emph{IP-dominate} $j$ if and only if $\overline
P_{SD}^{m,i} \le \overline P_{SD}^{m,j}$ and $R^{m,i} \ge R^{m,j}$,
then $j$ can be deleted from $\mathcal {Q}_m$. If $i$ is not
IP-dominate by any other $j \in \mathcal {Q}_m$, then $i$ is said to
be \emph{IP-efficient}. Let $\mathcal {Q}_m^e$ contains the
IP-feasible and IP-efficient indices of $\mathcal {Q}_m$.

we use dynamic programming to solve above problem [10]. In this
method, multiple-choice knapsack problem can be solved through
\emph{M} stages. While solving the \emph{L}-stage subproblem ($L \in
\mathcal {M}$), denoted as SP($L$), we only consider to use the
first $L$ DCs (denoted as $\mathcal {L} = \{1,\cdots,L\}$) for
transmission. The mathematical description of SP($L$) is
\begin{equation}
\begin{array}{l}
  \mathop {\max }\limits_{\left\{ {x^{l,q} } \right\}} \sum\limits_{l \in \mathcal {L}} {\sum\limits_{q \in \mathcal {Q}_l} {R^{l,q} x^{l,q} } }  ,  \vspace{3pt}
 \\
 {\rm{s}}{\rm{.t}}{\rm{.}}\;\left\{ \begin{array}{l}
 \sum\limits_{l \in \mathcal {L}} {\sum\limits_{q \in \mathcal {Q}_l} {\overline P_{SD}^{l,q}   x^{l,q} } }  \le P_{\max } ,  \vspace{3pt}
 \\
 \overline P_{SD}^{l,q}  \le P_{{\rm{max - s}}} (l),\;l \in \mathcal {L},\;q \in \mathcal {Q}_l,  \vspace{3pt}
 \\
 \sum\limits_{q \in \mathcal {Q}_l} {x^{l,q}  \le 1,\;} l \in \mathcal {L},  \vspace{3pt}
 \\
 x^{l,q}  \in \{ 0,\;1\} ,\;l \in \mathcal {L},\;q \in \mathcal {Q}_l.  \vspace{3pt}
 \end{array} \right. \\
 \end{array}
\end{equation}
A solution to SP($L$) can be uniquely characterized by a vector
${\bf{g}}$ with $g[l] \in \mathcal {Q}_l $ and $x^{l,g[l]}=1$, $l
\in \mathcal {L}$. Let $X_L^0$ denote a subset of the set of
solutions of SP($L$). For each $x\in X_L^0$, $\left( {R\left( x
\right),P\left( x \right)} \right)$ is said to be a DP-state, which
can be calculated as
\begin{equation}
\left( {R\left( x \right),P\left( x \right)} \right) = \left(
{\sum\limits_{l \in \mathcal {L}} {R^{l,g[l]} } ,\sum\limits_{l \in
L} {\overline P_{SD}^{^{l,g[l]} } } } \right).
\end{equation}
If $P\left( x \right) > P_{{\rm{max}}}$, the partial solution $x$ us
said to be \emph{DP-infeasible} and can be deleted from $X_L^0$. Let
$X_L^f$ denote a subset of the set of DP-feasible solutions of
SP($L$). If for $\forall x^d, x^e \in X_L^f$, which satisfy
\begin{equation}
P\left( {x^e } \right) \le P\left( {x^d }
\right)\;\;{\rm{and}}\;\;R\left( {x^e } \right) \ge R\left( {x^d }
\right),
\end{equation}
the partial solution $x^d$ is said to be DP-dominated by $x^e$ and
can be deleted from $X_L^f$. If $x^e$ is not DP-dominated by any
other element of $X_L^f$, $x^e$ is said to be DP-efficient. Let
$X_L^e$ denote the set of DP-feasible and DP-efficient solutions of
SP($L$). After solving $M$ subproblems, we can obtain the final
power allocation ${\bf{P}}_{SD}  = \left[ {P_{SD}^1 , \cdots
,P_{SD}^{M} } \right]$. In summary, the joint power-channel
allocation algorithm is described as the dynamic programming
algorithm shown in Fig. 3. \vspace{-5pt}

\section{Performance Analysis}
\label{Performance_Analysis}

Considering a node pair with source $S$ and destination $D$ with $M$
common available channels. The channel gains are $\left\{ {h_{SD}^1
, \cdots ,h_{SD}^M } \right\}$. The sets of available data
transmission rate and corresponding SNR threshold are the same as
those described in section~\ref{section_system_model}. Moreover, the
corresponding set of transmission radius on the $m$th DC is denoted
by $r^m = \left\{ {r_1^m , r_2^m, \cdots ,r_Q^m } \right\}$, where
$r_1^m
 > r_2^m >  \cdots  > r_Q^m$. We only consider the path loss component.

\begin{figure}
{\footnotesize \vspace{-3pt} \centerline{\begin{tabular}{p{8.5cm}}
\hrule \vspace{5pt} \textbf{~Dynamic programming algorithm: Joint
power-channel allocation.}\vspace{-3pt}
\\\hrule
%\hline
%\vspace*{-30pt}\\
\end{tabular}}
\begin{tabular}{p{0.3cm} p{7.7cm}}
~1)  &Initialization: $L=1$, $X_1^0 = N_1^e$;\\
~2)  &DP-dominance:\\
 &i) ~Construct $X_L^f$ by eliminating all DP-infeasible elements
of $X_L^0$;\vspace{1pt}\\
 &ii) Construct $X_L^e$ by eliminating all DP-infeasible elements
of $X_L^f$;\\
~3)  &If $L=M$, stop;\\
&Otherwise, $L=L+1$, construct $X_L^0 = X_{L-1}^e \times N_L^e$ and
goto 2).
\end{tabular}\vspace*{-5pt}
\centerline{\begin{tabular}{p{8.5cm}} \hrule
%\hline
%\vspace*{-30pt}\\
\end{tabular}}
}\vspace{-15pt} \caption{Dynamic programming algorithm for joint
power-channel allocation.}\vspace{-20pt}
\end{figure}

\subsection{Data Transmission Rate Analysis}

Suppose that node $S$ transmits on the $m$th DC with rate $R_q$.
When the interference caused by other nodes is absence, the SNR of
node $D$ which is apart from $S$ with $r_q^m$ can be expressed as
\begin{equation}
SNR_q \! =\! \frac{{P_r \left( {r_q ^m} \right)}}{{P_n }} \!=\!
\frac{{G_t \left( {f_m } \right)G_r \left( {f_m } \right)h_t^2 h_r^2
}}{{\left (r_q^m \right )^4  \cdot L \cdot P_n }}P_{\max },
\end{equation}
where $P_n$ is the noise power, and $f_m$ is the central frequency.
While considering the interference caused by other nodes, if the
distance between $S$ and $D$ is $d_{SD}$ and the transmit power on
the $m$th channel is $P_t^m$, then the SINR at $D$, denoted by
$SINR^m(d_{SD})$, is
\begin{equation}
SINR^m \left( {d_{SD} } \right)  = \frac{{G_t \left( {f_m }
\right)G_r \left( {f_m } \right)h_t^2 h_r^2 }}{{d_{SD}^4  \cdot L
\cdot \left( {P_n  + P_{\inf }^m } \right)}}P_t^m,
\end{equation}
where $P_r^m \left( {d_{SD} } \right)$ is the receiving power, and
$P_{inf}^m$ is the interference power on the $m$th channel detected
by $D$. If $S$ can use rate $R_q$ for transmission on the $m$th DC,
the inequation $SINR^m(d_{SD})  \ge SNR_q$ must be held.
Substituting (19) and (20) into the inequation, we can obtain the
relationship between $P_t^m$ and $P_{max}$, shown as
\begin{equation}
P_t^m  \ge \left( {\frac{{d_{SD} }}{{r_{q}^m }}} \right)^4
\frac{{P_n + P_{\inf }^m }}{{P_n }}P_{\max }.
\end{equation}
Suppose that the data transmission rate used by $S$ on the $m$th
channel is $R_{q(m)}$ ($m \in \mathcal {M}$ and $q\left( m \right)
\in \mathcal {Q}_m$), then the inequation, which is shown as
\begin{equation}
\sum\limits_{m \in \mathcal {M}} {P_t^m }  = \sum\limits_{m \in
\mathcal {M}} {\left( {\frac{{d_{SD} }}{{r_{q(m)}^m }}} \right)^4
\frac{{P_n  + P_{\inf }^m }}{{P_n }}} P_{\max }  \le P_{\max },
\end{equation}
must be satisfied. From the above inequation, we can get the
condition that $d_{SD}$ must hold, which is given by
\begin{equation}
d_{SD}  \le \left[ {\sum\limits_{m \in \mathcal {M}} {\left(
{\frac{1}{{r_{q(m)}^m }}} \right)^4 \frac{{P_n  + P_{\inf }^m
}}{{P_n }}} } \right]^{ - \frac{1}{4}}.
\end{equation}
If we denote the set of transmission radius on the CCC as $r =
\left\{ {r_1 , \cdots ,r_Q } \right\}$, the relationship between
$r_q^m$ and $r_q$ is
\begin{equation}
r_q ^m = \left( {{{f_m } \mathord{\left/
 {\vphantom {{f_m } {f_0 }}} \right.
 \kern-\nulldelimiterspace} {f_0 }}} \right)r_q.
\end{equation}
From (24) into (23), the constraint for $d_{SD}$ can be written as
\begin{equation}
d_{SD}  \le \left[ {\sum\limits_{m \in \mathcal {M}} {\left(
{\frac{1}{{r_q }} \cdot \frac{{f_0 }}{{f_m }}} \right)^4 \frac{{P_n
+ P_{\inf }^m }}{{P_n }}} } \right]^{ - \frac{1}{4}}.
\end{equation}
The total data transmission rate $R_{SD}$ can be calculated as:
\begin{equation}
R_{SD}  = \sum\limits_{m \in \mathcal {M}} {R_{q(m)} }.
\end{equation}
From (25) and (26), we get the relationship between the distance and
the total data transmission rate. If the transmission rate on each
channel is given and the interference is measured, the required
distance between source and destination can be calculated. Since the
number of total transmission rate is $Q^M$, the number of
corresponding distance is also $Q^M$, which provide more
adaptability for transmission rate.

\subsection{Throughput Analysis}

Suppose that the power allocation of node $S$ and $D$ is
${\bf{P}}_{SD} = \left[ {P_{SD}^1 , \cdots ,P_{SD}^M } \right]$. For
the $m$th ($m \in \mathcal {M}$) channel, the probability that $S$
can use rate $R_q$ ($q \in \{1,\cdots,Q-1 \}$) for transmission on
the channel is
\begin{equation}
\Pr \left\{ {R_{SD}^m \!\! = \!\!R_q } \right\} \!=\!\! \Pr
\!\left\{ {SNR_q\! \le \!SINR_{SD}^m \! < \!SNR_{q + 1} } \!
\right\}
\end{equation}
where $R_{SD}^m$ is the transmission rate on the $m$th channel and
$SINR_{SD}^m$ is the SINR of node $D$ on the channel. Substituting
(19) and (20) into the above equation, we can obtain
\begin{equation}
\Pr \{ R_{SD}^m  = R_q \}  = \Pr \left\{ {\Gamma ^m r_{q + 1}^m  \le
d_{SD}  < \Gamma ^m r_q^m } \right\},
\end{equation}
where $\Gamma ^m  = \left[ {{{P_{SD}^m P_n } \mathord{\left/
 {\vphantom {{P_{SD}^m P_n } {P_{\max } \left( {P_n  + P_{\inf }^m } \right)}}} \right.
 \kern-\nulldelimiterspace} {P_{\max } \left( {P_n  + P_{\inf }^m } \right)}}} \right]^{1/4}
 $.
Through similar approach, we can obtain the probabilities that $S$
uses rate $R_Q$ for transmission and gives up using this channel,
respectively, which are shown as
\begin{eqnarray}
 \Pr \{ R_{SD}^m  = R_Q \}  &= &\Pr \{ SINR_{SD}^m  \ge SNR_Q \} \nonumber\\
                            &= &\Pr \{ d_{SD}  \le \Gamma ^m r_Q^m
                            \},
\end{eqnarray}
\begin{eqnarray}
 \Pr \{ R_{SD}^m  = 0\}  &= \Pr \{ SINR_{SD}^m  < SNR_1 \}  \nonumber\\
                         &= \Pr \{ d_{SD}  > \Gamma ^m r_1^m \}.
\end{eqnarray}
If the distribution of nodes is given, we can obtain the probability
density function (pdf) of the distance $d_{SD}$ and calculate the
probabilities shown in (28)-(30). For example, if the nodes are
uniformly distributed, then the pdf of $d_{SD}$ is
$f(d_{SD})=2d_{SD}$. Therefore, (28)-(30) become
\begin{eqnarray}
 \Pr \{ R_{SD}^m  = R_q \}  &= \int_{\Gamma ^m r_{q + 1}^m }^{\Gamma ^m r_q^m } {f(d_{SD} )d(d_{SD} )}  \nonumber\\
                            &= (\Gamma ^m )^2 [(r_q^m )^2  - (r_{q + 1}^m )^2
                            ],
\end{eqnarray}
\begin{equation}
\Pr \{ R_{SD}^m  = R_Q \}\!  = \!\!\int_0^{\Gamma ^m r_Q^m }
{f(d_{SD} )d(d_{SD} )} \! = \!(\Gamma ^m r_Q^m )^2,
\end{equation}
\begin{equation}
\Pr \{ R_{SD}^m  = 0\} \! = \!1\! -\! \sum\limits_{q = 1}^Q {\Pr \{
R_{SD}^m = R_q \} } \! = \!1 \!-\! (\Gamma ^m r_1^m )^2.
\end{equation}
Therefore, the expected average transmission rate is
\begin{equation}
 \mathbb{E}\left\{ {R_{SD} } \right\} = \sum\limits_{m = 1}^M {\sum\limits_{q = 0}^Q {\Pr \left\{ {R_{SD}^m  = R_q } \right\}  R_q }
 }.
\end{equation}
Using $\mathbb{E}\left\{ {R_{SD} } \right\}$ instead of $R_{SD}$ in
(11), we can get the expected average number of data packets
$\mathbb{E}\left\{ {N_{SD} } \right\}$ that can be transmitted
during one transmission process. Then, using $\mathbb{E}\left\{
{N_{SD} } \right\}$ and $\mathbb{E}\left\{ {R_{SD} } \right\}$
instead of $N_{SD}$ and $R_{SD}$ in (10), the expected average
transmission time for one transmission process, denoted by
$\mathbb{E}\left\{ {T_{SD} } \right\}$ can be calculated.

Finally, we can get the expected average throughput between $S$ and
$D$:
\begin{equation}
\mathbb{E}\left\{ {\Psi_{SD} } \right\} = {{E\left[ {N_{SD} }
\right]  L_{\rm{{data} }}} } \mathord{\left/
 {\vphantom {{E\left[ {N_{SD} } \right] L_{\rm{{data} }}} {E\left[ {T_{SD} } \right]}}} \right.
 \kern-\nulldelimiterspace} {E\left[ {T_{SD} } \right]}.
\end{equation}

\section{Simulation Results}
\label{Simulation_Results}

Suppose each node supports three different data transmission rates
which are 2, 5.5, and 11Mbps, respectively. The corresponding
transmission radii of the CCC are 250, 200, and 100m, respectively.
The control packets are exchanged on the CCC with basic rate 2Mbps.
Each data packet contains 1000 Bytes.

Fig. 4 shows the data transmission rate gains of the MCD-MAC
protocol with different number of data channels over the data rate
with one data channel. From the result, we can find that the data
transmission rates with multiple data channels are significantly
larger than that with one data channel, and the obtained gains are
the most notable when the distance between source and destination is
small. The reason is that the required transmitting power is small
with the small distance, therefore, nodes can utilize all the data
channels under the total power limitation. The simulation result
also shows, no matter how many channels are available, the obtained
data rate gains decrease as the distance between the two nodes
increases. The reason is that the propagation loss will become
larger with longer distance, in order to finish their transmission,
the source has to decrease its transmission rate to guarantee the
SINR at the destination. Although the data transmission rate with
multiple data channels equals that with one channel under extremely
long distance, the multi-channel diversity can still significantly
improve the data transmission rate in most of the distance range.

Fig. 5 shows the average node throughput of the MCD-MAC protocol as
a function of number of available data channels under different
interference power. From the result we can find, no matter how many
channels can be used, the average node throughput decreases as the
interference power increases. This is because larger transmit power
is needed to guarantee that the SINR at the destination is larger
than the threshold when interference exists, which causes the source
node has to use less data channels or choose lower transmission rate
because of the limited transmit power resources. However, given the
interference power, the average node throughput with multiple data
channels is larger than that with one channel. In particular, the
more data channels are used, the more obvious the improvement is,
which shows the advantage of multi-channel diversity.

\begin{figure}[t]
\begin{center}
\includegraphics[width=8cm]{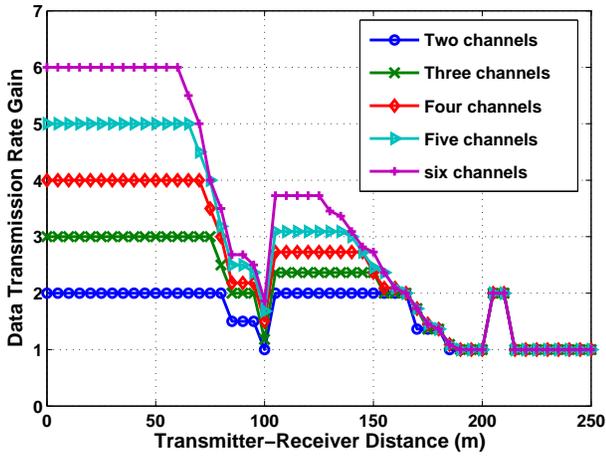}
\end{center}
\caption{Normalized data transmission rate gains.} \label{fig:4}
\end{figure}

\begin{figure}[t]
\begin{center}
\includegraphics[width=8cm]{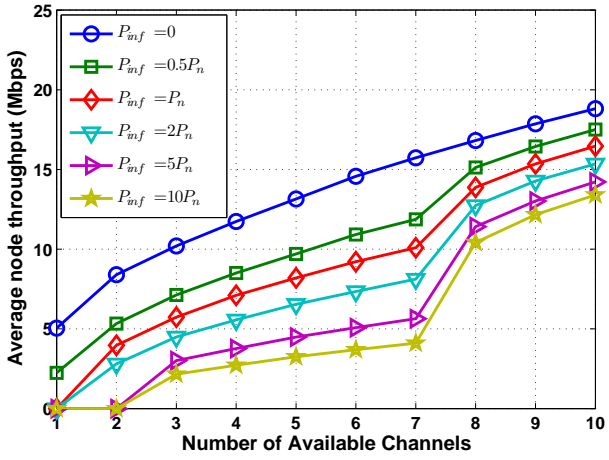}
\end{center}
\caption{Average node throughput of MCD-MAC under different
interference power.} \label{fig:5}
\end{figure}

Fig. 6 shows the average network throughput of MCD-MAC, OMMAC [4]
with power limitation and MOAR [2] as a function of number of flows.
In our simulation, nodes are uniformly distributed in a circular
area with a diameter of 250m, and any node randomly chooses one of
its neighbors as its destination. The network contains six data
channels. In order to simulate the PUs' activities, the probability
that each channel is occupied by PUs in each CS is 0.5, which means
half of channels can be used statistically. The simulation result
shows that the average network throughput of the MCD-MAC protocol
obviously exceeds those of MOAR and OMMAC with total power
limitation. The main reason is that the MCD-MAC protocol uses
multi-channel diversity, which can help nodes to efficiently and
fully utilize available opportunistic DCs for data transmission.
Moreover, the most suitable number of channels that they will use
and the corresponding power allocation can be dynamically adjusted
by joint channel/power allocation according to the distance between
source and destination as well as the interference caused by their
neighbor nodes. Furthermore, because of the power control brought by
the multi-channel diversity, the mutual interference among neighbor
nodes is reduced, and the space reuse efficiency is improved, which
are benefit for the throughput improvement.

\begin{figure}[t]
\begin{center}
\includegraphics[width=8cm]{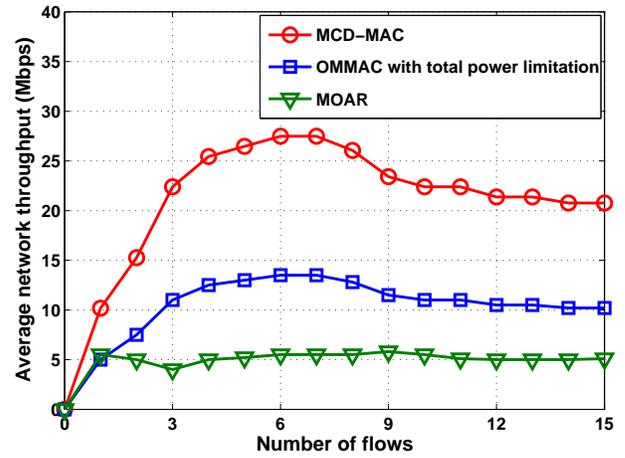}
\end{center}
\caption{Throughput of MCD-MAC, MOAR, and OMMAC.} \label{fig:6}
\end{figure}

\section{Conclusion}
\label{Conclusion}

In this paper, we first propose the multi-channel diversity (MCD),
which allows each secondary node to utilize multiple channels
simultaneously with only one radio per node and upperbounded
transmit power. Then, using the proposed MCD, we develop a novel MAC
protocol, named MCD-MAC. The protocol can efficiently utilize
available channel resources through joint power-channel allocation.
Particulary, we convert the joint power-channel allocation to the
multiple-choice knapsack problem such that the optimal transmission
strategy can be obtained through dynamic programming. In addition,
we also design a mechanism to guarantee the transmission-time
fairness for different node pairs. Simulation results show that our
proposed MCD-MAC protocol outperforms the existing protocols.

\end{document}